\documentclass[10pt, conference]{IEEEtran}
\PassOptionsToPackage{caption=false,font=footnotesize}{subfig}
\usepackage{amsmath,amsfonts,amssymb,algpseudocode}
\usepackage{algorithmicx}
\usepackage{algorithm2e}
\usepackage{array}
\usepackage{booktabs}
\usepackage{textcomp}
\usepackage{stfloats}
\usepackage{url}
\usepackage{verbatim}
\usepackage{graphicx}
\usepackage{comment}
\usepackage{enumitem}

\usepackage{fontspec}

\usepackage{subfig}

\usepackage{siunitx}
\DeclareSIUnit{\Mbps}{Mbps}
\usepackage[
bibstyle=ieee,
citestyle=numeric-comp,
maxnames=1,
minnames=1,
backend=biber,
]{biblatex}
\usepackage{balance}
\usepackage[hidelinks]{hyperref}
\usepackage{cleveref}

\usepackage{xcolor}
\definecolor{aqua}{RGB}{63, 152, 189}
\definecolor{mybgcolor}{HTML}{b2d6e5}
\definecolor{slide1Freq}{HTML}{3e98bd}
\definecolor{bkoffcol}{HTML}{c69404}
\definecolor{labelfont}{HTML}{90288d}
\definecolor{stasenddata}{HTML}{e2f0d9}
\definecolor{algoap}{HTML}{c00000}
\definecolor{phaseBSR}{HTML}{D9EAD3}
\definecolor{phaseUL}{HTML}{D9D2E9}
\definecolor{phaseACK}{HTML}{EAD1DC}
\definecolor{apBcast}{HTML}{BDD7EE}

\DeclareSIUnit{\dBm}{dBm}

\usepackage[abbreviations,nomain,nonumberlist]{glossaries-extra}
\setabbreviationstyle{long-short}   

\newabbreviation{aiot}{A-IoT}{Ambient Internet of Things}
\newabbreviation{am}{AM}{Acknowledged Mode}
\newabbreviation{ams}{AMC}{Adaptive Modulation and Coding}
\newabbreviation{arq}{ARQ}{Automatic Repeat reQuest}

\newabbreviation{bs}{BS}{Base Station}
\newabbreviation{bsr}{BSR}{Buffer Status Report}

\newabbreviation{cdrx}{cDRX}{Connected-Mode Discontinuous Reception}
\newabbreviation{ce}{CE}{Coverage Enhancement}
\newabbreviation{crnti}{C-RNTI}{Cell Radio Network Temporary Identifier}

\newabbreviation{dci}{DCI}{Downlink Control Information}
\newabbreviation{drx}{DRX}{Discontinuous Reception}

\newabbreviation{edrx}{eDRX}{Extended Discontinuous Reception}
\newabbreviation{edt}{EDT}{Early Data Transmission}
\newabbreviation{enb}{eNB}{Evolved Node B}
\newabbreviation{epc}{EPC}{Evolved Packet Core}
\newabbreviation{ewma}{EWMA}{Exponentially Weighted Moving Average}

\newabbreviation{fug}{FUG}{Fast Uplink Grant}

\newabbreviation{gwus}{GWUS}{Group Wake-Up Signal}

\newabbreviation{hsr}{HSR}{Hybrid Scheduling Request}

\newabbreviation{iot}{IoT}{Internet of Things}

\newabbreviation{lte}{LTE}{Long Term Evolution}

\newabbreviation{mac}{MAC}{Medium Access Control}
\newabbreviation{mcs}{MCS}{Modulation and Coding Scheme}
\newabbreviation{mib}{MIB}{Master Information Block}
\newabbreviation{mme}{MME}{Mobility Management Entity}
\newabbreviation{mtc}{MTC}{Machine Type Communications}

\newabbreviation{nb-iot}{NB-IoT}{Narrowband Internet of Things}
\newabbreviation{npdcch}{NPDCCH}{Narrowband Physical Downlink Control Channel}
\newabbreviation{npdsch}{NPDSCH}{Narrowband Physical Downlink Shared Channel}
\newabbreviation{nprach}{NPRACH}{Narrowband Physical Random Access Channel}
\newabbreviation{npss}{NPSS}{Narrowband Primary Synchronization Signal}
\newabbreviation{npusch}{NPUSCH}{Narrowband Physical Uplink Shared Channel}
\newabbreviation{nsss}{NSSS}{Narrowband Secondary Synchronization Signal}

\newabbreviation{pdcp}{PDCP}{Packet Data Convergence Protocol}
\newabbreviation{pdr}{PDR}{Packet Delivery Ratio}
\newabbreviation{pdu}{PDU}{Protocol Data Unit}
\newabbreviation{pgw}{PGW}{Packet Data Network Gateway}
\newabbreviation{psm}{PSM}{Power Saving Mode}
\newabbreviation{pur}{PUR}{Preconfigured Uplink Resources}

\newabbreviation{ra}{RA}{Random Access}
\newabbreviation{rai}{RAI}{Release Assistance Indication}
\newabbreviation{rar}{RAR}{Random Access Response}
\newabbreviation{rlc}{RLC}{Radio Link Control}
\newabbreviation{rnti}{RNTI}{Radio Network Temporary Identifier}
\newabbreviation{rrc}{RRC}{Radio Resource Control}

\newabbreviation{sgw}{SGW}{Serving Gateway}
\newabbreviation{sps}{SPS}{Semi-Persistent Scheduling}
\newabbreviation{sr}{SR}{Scheduling Request}

\newabbreviation{tau}{TAU}{Tracking Area Update}

\newabbreviation{udp}{UDP}{User Datagram Protocol}
\newabbreviation{ue}{UE}{User Equipment}
\newabbreviation{ul}{UL}{Uplink}
\newabbreviation{uss}{USS}{User-Specific Search Space}

\newabbreviation{3gpp}{3GPP}{Third Generation Partnership Project}

\usepackage{tikz}
\usetikzlibrary{arrows.meta, positioning, calc, backgrounds, shapes.symbols,
shapes.geometric, fit, decorations.pathreplacing}

\usepackage{silence}
\graphicspath{{./images}}

\usepackage[pro]{fontawesome6}

\usepackage{nicefrac}

\begin{document}

\title{
    System-Level Evaluation of Uplink Access Schemes for Ambient IoT
}

\author{\IEEEauthorblockN{Douglas Dziedzorm Agbeve,  Henrique Duarte Moura, Andrey Belogaev, Jeroen Famaey}
\IEEEauthorblockA{
    \textit{IDLab, University of Antwerp – imec, Belgium}\\
\{douglas.agbeve, henrique.duartemoura, andrei.belogaev, jeroen.famaey\}@uantwerpen.be}
}

\maketitle

\begin{abstract}
    \gls{aiot} devices operate under stringent energy constraints, relying on
    harvested power in the \unit{\micro\watt}-to-\unit{\milli\watt} range and
    limited energy storage. The \gls{aiot} air interface for active devices is
    still under development; we therefore use \gls{nb-iot} as an implementable
    reference for evaluating autonomous \gls{ul} access under harvested energy.
    Under these conditions, the standard \gls{nb-iot} contention-based \gls{ra}
    procedure incurs substantial per-packet overhead, making \gls{ul} access in
    dense A-IoT networks increasingly energy-expensive. Context-retaining
    \gls{ul} access schemes reduce this overhead, but their energy–latency
    trade-offs are insufficiently studied in literature under realistic
    harvesting and protocol constraints. To address this gap, we extend the
    LENA-NB \texttt{ns-3} module with solar energy harvesting, supercapacitor
    storage, and a hysteretic brown-out mechanism, and propose \gls{hsr}, a
    hybrid \gls{ul} access scheme combining reserved request occasions with
    opportunistic contention. We compare \gls{ra}, \gls{sr}, \gls{hsr} and an
    Idealized \gls{fug} serving as a performance bound. Results show that
    \gls{hsr} reduces mean packet delay by up to \qty{69}{\percent} and \gls{sr}
    reduces energy per delivered bit by up to \qty{24}{\percent}, both relative
    to \gls{ra}.

\end{abstract}

\glsresetall

\begin{IEEEkeywords}
    ambient IoT, NB-IoT, energy harvesting, uplink access, scheduling request,
    fast uplink grant, ns-3.
\end{IEEEkeywords}

\section{Introduction}

\looseness=-1

The \gls{3gpp} is standardizing \gls{aiot} as a new class of cellular \gls{iot}
devices targeting battery-less operation through ambient energy
harvesting~\cite{3gpp_tr38848}. Release~19 specifies an air
interface for passive devices that backscatter a carrier provided by the
network~\cite{ts38291,ts38391}. Release~20 extends the framework to active
devices, including types~2b and~C, which generate their own transmissions and
therefore do not require an externally supplied carrier. The air interface for
these active device types remains unspecified in the current Release~20 work
item~\cite{rp260826}. We therefore use \gls{nb-iot}, an already deployed
low-power cellular \gls{iot} technology, as an implementable reference. Unlike
conventional battery-powered \gls{nb-iot} \glspl{ue}, \gls{aiot} devices operate
under extremely constrained energy budgets of microwatts to milliwatts. The
harvested energy is commonly buffered in supercapacitors, which have lower
energy density and higher self-discharge than batteries, making device operation
dependent on instantaneous energy availability. Furthermore, their traffic is
inherently sparse, consisting of short and infrequent \gls{ul} reports rather
than sustained data flows. Under these conditions, control-plane signaling can
dominate the overall energy consumption, while insufficient harvested energy may
trigger brown-out events that interrupt both communication and application
execution~\cite{butt_aiot_2024}.

The standard \gls{nb-iot} \gls{ul} access procedure introduces substantial
per-packet overhead through the four-step contention-based \gls{ra} process,
requiring preamble transmission, random-access response, contention
resolution, and \gls{ul} resource allocation before application data delivery
~\cite{Michelinakis_2020}. Context-retaining schemes, such as \gls{sps},
\gls{pur}, and \gls{gwus}~\cite{hoglund_pur_2020,medina_acosta_rel17_2022},
as well as predictive \gls{fug} scheduling~\cite{ali_fug_2019}, mitigate this
repeated overhead by allowing lightweight data requests or proactive resource
allocation. However, they introduce additional listening, state-retention,
and resource-partitioning costs. Moreover, reserved request opportunities
must be shared among devices, causing latency to increase with the device
density.

The suitability of context retention under active \gls{aiot} operation depends
jointly on traffic characteristics, harvesting dynamics, and network load, and
has not been quantified. Similar access and context-management choices are still
open in the standardization process. For types~2b and~C devices,
current agreements allow either contention-based access carrying data or a
preamble-only request transmitted on resources shared by multiple
devices~\cite{ran1_126}. Whether a device may retain its temporary identifier
and thereby avoid repeated random access also remains undecided~\cite{ran2_135}.
Existing work either assumes periodic traffic (\gls{sps}, \gls{pur}), accurately
predictable traffic (\gls{fug}), or unconstrained energy availability. Hybrid
approaches combining reserved and opportunistic request opportunities remain
unexplored.

To address this gap, we extend the LENA-NB \texttt{ns-3} framework
~\cite{Jorke2022} with a harvest-aware \gls{aiot} energy front-end comprising
a solar energy harvester, a supercapacitor-based storage model, and a
hysteretic brown-out mechanism that couples energy availability with both
traffic generation and radio operation. Building upon this framework, we
implement: (i) context-retaining \gls{ul} access through dedicated \glspl{sr} on
reserved \gls{nprach} resources and (ii) propose \gls{hsr}, a hybrid scheme that
combines reserved requests with energy-aware opportunistic contention.
\Gls{sr} represents preamble-only requesting when request resources are
dedicated, \gls{hsr} extends this approach to request resources shared among
devices, and \gls{ra} with \gls{edt} represents contention-based access that
carries data within the access procedure~\cite{ran1_126,ran2_135}. This enables
the first system-level comparison of \gls{ra}, \gls{sr}, and \gls{hsr}, under
identical harvesting, traffic, and propagation conditions, with an Idealized
\gls{fug} configuration bounding the achievable latency and energy.

Sections~\ref{sec:uplink_mac}--\ref{sec:relwk} present the access schemes
and related work, Sections~\ref{sec:system_model}--\ref{sec:perf_eval} detail
the system model, simulation platform, and results, and
Section~\ref{sec:conclusion} concludes the paper.

\section{Uplink Access Schemes}
\label{sec:uplink_mac}

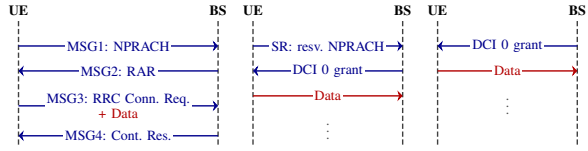
\begin{figure}[tb]
  \centering
  \scalebox{0.66}{
  \begin{tikzpicture}[
    >={Stealth[length=3pt, width=6pt]},
    font=\footnotesize,
    control/.style  = {->, thick, color=blue!55!black},
    data/.style     = {->, thick, color=red!70!black, line width=0.8pt},
    lifeline/.style = {-, thick, gray!50!black, dash pattern=on 3pt off 1.5pt},
    nodelabel/.style = {font=\bfseries\footnotesize, anchor=south},
    msglabel/.style  = {midway, font=\footnotesize, fill=white, inner sep=1pt},
  ]

  \coordinate (UE1) at (0,   0);
  \coordinate (BS1) at (4.0, 0);
  \node[nodelabel] at (UE1) {\glsxtrshort{ue}};
  \node[nodelabel] at (BS1) {\glsxtrshort{bs}};
  \draw[lifeline] (UE1) -- ++(0, -2.5);
  \draw[lifeline] (BS1) -- ++(0, -2.5);

  \draw[control] ([yshift=-0.5cm]UE1) -- ([yshift=-0.5cm]BS1)
      node[msglabel] {MSG1: NPRACH};
  \draw[control] ([yshift=-1.0cm]BS1) -- ([yshift=-1.0cm]UE1)
      node[msglabel] {MSG2: RAR};
  \draw[control] ([yshift=-1.7cm]UE1) -- ([yshift=-1.7cm]BS1)
      node[msglabel,align=center] {MSG3: RRC Conn.\ Req.\\
      \textcolor{red!70!black}{+ Data}};
  \draw[control] ([yshift=-2.3cm]BS1) -- ([yshift=-2.3cm]UE1)
      node[msglabel] {MSG4: Cont.\ Res.};

  \coordinate (UE2) at (4.7, 0);
  \coordinate (BS2) at (7.7, 0);
  \node[nodelabel] at (UE2) {\glsxtrshort{ue}};
  \node[nodelabel] at (BS2) {\glsxtrshort{bs}};
  \draw[lifeline] (UE2) -- ++(0, -2.5);
  \draw[lifeline] (BS2) -- ++(0, -2.5);

  \draw[control] ([yshift=-0.5cm]UE2) -- ([yshift=-0.5cm]BS2)
      node[msglabel] {\glsxtrshort{sr}: resv.\ NPRACH};
  \draw[control] ([yshift=-1.0cm]BS2) -- ([yshift=-1.0cm]UE2)
      node[msglabel] {DCI~0 grant};
  \draw[data]    ([yshift=-1.5cm]UE2) -- ([yshift=-1.5cm]BS2)
      node[msglabel] {Data};
  \node[font=\footnotesize, color=gray!60!black]
       at ($(UE2)!0.5!(BS2) + (0, -2.1)$) {$\vdots$};

  \coordinate (UE3) at (8.4,  0);
  \coordinate (BS3) at (11.2, 0);
  \node[nodelabel] at (UE3) {\glsxtrshort{ue}};
  \node[nodelabel] at (BS3) {\glsxtrshort{bs}};
  \draw[lifeline] (UE3) -- ++(0, -2.5);
  \draw[lifeline] (BS3) -- ++(0, -2.5);

  \draw[control] ([yshift=-0.5cm]BS3) -- ([yshift=-0.5cm]UE3)
      node[msglabel] {DCI~0 grant};
  \draw[data]    ([yshift=-1.0cm]UE3) -- ([yshift=-1.0cm]BS3)
      node[msglabel] {Data};
  \node[font=\footnotesize, color=gray!60!black]
       at ($(UE3)!0.5!(BS3) + (0, -1.6)$) {$\vdots$};

  \end{tikzpicture} }
  \caption{\gls{ul} access schemes: \gls{ra} with \gls{edt} (left), \gls{sr} (center), and
           \gls{fug} (right).}
  \label{fig:ra_fug_scheme}
  \vspace{-.6cm}
\end{figure}
This section presents the \gls{ul} channel access schemes considered in this
work: \gls{ra}, \gls{fug}, \gls{sr}, and \gls{hsr}. The signaling procedures of
the schemes are illustrated in Fig.~\ref{fig:ra_fug_scheme}.

\subsection{Random access (RA) with \glsfmtshort{edt}}
Under \gls{ra}, each \gls{ul} transmission is preceded by the standard two- to
four-step contention-based access procedure defined in 3GPP TS~36.321. A
\gls{ue} returning from  \gls{psm} or \gls{edrx} first transmits an MSG1
preamble over the configured \gls{nprach} resources. After successful detection,
the \gls{bs} responds with MSG2 containing a temporary identifier and an
\gls{ul} grant for MSG3 transmission, which carries the \gls{rrc} connection
request and, in an \gls{edt} configuration, \gls{ul} payload. The
procedure is completed through contention resolution (MSG4). If multiple
\glspl{ue} select the same preamble, the affected devices may perform back-off
and retransmission up to the configured attempt limit, after
which the packet is discarded. Following transmission completion, the \gls{ue}
returns to \gls{edrx} for duration $T_{3324}$ before transitioning back to
\gls{psm} until the next reporting event or expiration of timer $T_{3412}$.

From a resource-allocation perspective, each RA attempt consumes a sequence of
\gls{nprach}, \gls{npdcch}, and \gls{npusch} resources whose availability is
bounded by the cell broadcast configuration. \gls{nprach} opportunities are
configured with periodicity
$T_\mathrm{NPRACH}\in[40\,\si{\milli\second},\,2.56\,\si{\second}]$, with each
occasion reserving $P_\mathrm{sc}$ contiguous subcarriers for \gls{ra}
transmission. A single-tone Format-0 preamble occupies $N_\mathrm{rep}\cdot
5.6\,\si{\milli\second}$, where the repetition factor
$N_\mathrm{rep}\in\{1,\ldots,128\}$ is selected according to the coverage
enhancement level. Following MSG1 reception, the \gls{bs} returns MSG2 within
the configured random-access response window $W_\mathrm{RAR}$, while the uplink
scheduling delay for MSG3 is determined by the parameter
$k_2\in\{12,16,32,64\}\,\si{\milli\second}$. The resulting access delay can
therefore be approximated as
$
\tau_\mathrm{RA} \;\le\;
T_\mathrm{NPRACH} + W_\mathrm{RAR} + k_2 +
T_\mathrm{NPUSCH} + T_\mathrm{CR},
\label{eq:tau_ra}
$
where $T_\mathrm{NPUSCH}$ denotes the MSG3 transmission duration and
$T_\mathrm{CR}$ represents the contention-resolution turnaround time.

\subsection{Fast Uplink Grant (\glsfmtshort{fug})}

\gls{fug} requires the full \gls{ra} procedure only during the initial \gls{ul}
transmission of each \gls{ue}. Once attached, both the \gls{ue} and the \gls{bs}
retain the transmission context established during the initial \gls{ra}
procedure, including the \gls{crnti}, timing-advance information, and all state
information required to resume scheduled \gls{ul} transmissions without
re-initiating \gls{ra} across sleep cycles. For stationary \glspl{ue}, where
the propagation delay remains effectively constant, the stored context remains
valid across power-saving periods (i.e., \gls{psm} and \gls{edrx}), and the
\gls{bs} proactively schedules \gls{ul} transmissions via \gls{dci}~0 grants
over the \gls{uss}, whose periodicity is
$T_\mathrm{USS}=R_\mathrm{max}\cdot G$ subframes, where $R_\mathrm{max}$
denotes the maximum number of \gls{npdcch} repetitions and
$G\in\{1.5,2,4,\ldots,64\}$ the scheduling-gap factor. After decoding the
grant, the \gls{ue} transmits the corresponding \gls{npusch} burst following a
scheduling delay $k_0\in\{8,16,32,64\}\,\si{\milli\second}$, so the uplink
delay is bounded as
$\tau_\mathrm{FUG}\le T_\mathrm{USS} + k_0 + T_\mathrm{NPUSCH}$.

What remains is how the \gls{bs} decides when to push a grant: an
idealized variant, used here as a performance bound, learns of each buffer
arrival instantly at no signaling cost, whereas a deployable variant must
predict each \gls{ue}'s reporting instants from its observed traffic history.
The idealized variant therefore bounds both the achievable latency and the
achievable energy of request-based access.

\subsection{Scheduling Request (\glsfmtshort{sr}) and Hybrid
\glsfmtshort{sr} (\glsfmtshort{hsr})}

As in \gls{fug}, under \gls{sr} both the \gls{ue} and the \gls{bs} retain the
connection context across periods of inactivity and \gls{psm} intervals once the
initial \gls{ul} transmission has been completed. While \gls{sr} transmission on
dedicated \gls{nprach} resources is standardized for connected \glspl{ue} in
Rel-15, its use here as a resume trigger from \gls{psm} with retained context
extends the standard; a functionally equivalent realization is introduced in
Rel-16 through \gls{pur}. Following the initial connection setup, the \gls{bs}
reserves a specific \gls{nprach} occasion and sub-carrier for each \gls{ue} and
associates these resources with its retained connection context. The resource
reservation is valid for a period of $T_\mathrm{res}$ and is renewed upon each
access and during the periodic \gls{tau} procedure. If the reservation expires,
the \gls{ue} loses its reserved resources and must re-establish the connection
through \gls{ra}. When data arrives, the \gls{ue} wakes, re-synchronizes via
\glsfmtshort{npss}, \glsfmtshort{nsss} and \glsfmtshort{mib}, and requests an
\gls{ul} grant by transmitting a preamble during its pre-assigned \gls{nprach}
occasion on its reserved sub-carrier. Upon detecting the preamble, the \gls{bs}
allocates an uplink grant via \gls{dci}~0. Since only a fraction of the
\gls{nprach} sub-carriers can be reserved, the available resource elements are
time-multiplexed among the \glspl{ue} in a round-robin fashion. Consequently, a
given \gls{ue}'s reserved occasion recurs only every
$T_\mathrm{SR}=T_\mathrm{base}\cdot\big\lceil N/N_\mathrm{res}\big\rceil,$ where
$T_\mathrm{base}$ denotes the base \gls{nprach} occasion period, $N$ is the
number of \glspl{ue}, and $N_\mathrm{res}$ is the number of reserved
sub-carriers. With $N_\mathrm{res}{=}6$ and
$T_\mathrm{base}{=}\qty{80}{\milli\second}$, $T_\mathrm{SR}$ grows linearly with
$N$, exceeding one second beyond $N\approx 75$; at scale, request latency rather
than data transmission therefore dominates the overall uplink delay.

To mitigate this request latency, we propose a \emph{hybrid} scheme that
combines dedicated and contention-based request opportunities. Rather than
waiting for its reserved \gls{nprach} occasion, a \gls{ue} with pending data may
transmit a preamble on a randomly selected non-reserved sub-carrier during each
base \gls{nprach} occasion, provided that sufficient energy is available and the
remaining time before its assigned \gls{nprach} occasion is sufficient to
complete the \gls{ra} procedure. If the preamble transmission succeeds, the
\gls{ue} proceeds with the remaining \gls{ra} steps and transmits its pending
data. Otherwise, it retries at the next base \gls{nprach} occasion while the
timing constraint remains satisfied. Consequently, the request latency is
bounded by approximately $T_\mathrm{base}$ and $T_\mathrm{SR}$, corresponding to
the opportunistic and reserved request opportunities, respectively.

\section{Related work}
\label{sec:relwk}

The energy cost of contention-based \gls{ra} access in \gls{nb-iot} has been
studied both analytically and experimentally. Prior work has characterized
\gls{nb-iot} energy consumption across \gls{psm}, \gls{edrx}, and connected
states~\cite{sultania_2018_energy,nbiot_energy_diverse_2023}, while measurements
on commercial hardware identify \gls{nprach} transmission as a dominant
contributor to the per-packet energy budget for short \gls{ul}
payloads~\cite{Michelinakis_2020}. Other studies analyse the
\gls{nprach}--\gls{npusch} resource trade-off and its impact on
collision-limited operation~\cite{liu_nprach_three_ce_2020}. However, these
works assume an unconstrained energy supply and therefore do not consider
\gls{aiot} devices.

To reduce repeated access overhead, \gls{3gpp} introduced persistent \gls{ul}
schemes such as \gls{sps}, \gls{pur}, and
\gls{gwus}~\cite{TS36213,hoglund_pur_2020}. In parallel, the
\gls{fug} scheme was proposed~\cite{ali_fug_2019}, followed by predictive and
learning-based schedulers for massive \gls{mtc}, including multi-armed-bandit
allocation~\cite{ali_sleeping_mab_2019} and learning-based traffic-prediction
and activation models~\cite{eldeeb_hmm_2022}. Closer to \gls{nb-iot}, a persistent
periodic scheduler exploiting power-saving and suspend/resume procedures has
been proposed~\cite{Wu_persistent_periodic_2022}, while grant-free and
\gls{fug} access have been analytically benchmarked against
\gls{ra}~\cite{Moons_evaluating_fug_2024}. However, existing learning-based
approaches consider simplified system models with a focus on the MAC layer, whereas
\gls{nb-iot}-specific studies rely on fixed periodic
heuristics~\cite{Wu_persistent_periodic_2022} or analytical
benchmarking~\cite{Moons_evaluating_fug_2024}. Consequently, the performance
trade-offs of the \gls{ul} access schemes under energy-harvesting constraints
remain unexplored, and hybrid schemes combining persistent and opportunistic
access have not been investigated.

Open-source \texttt{ns-3} support for \gls{nb-iot} evolved from early
energy-aware LTE models~\cite{sultania_ns3_2019} to
LENA-NB~\cite{Jorke2022}, which implements Rel-13/14 features including
cross-subframe scheduling. However, existing implementations support neither
energy harvesting operation, supercapacitor storage, and brown-out behavior,
nor persistent and \gls{fug} \gls{ul} scheduling, preventing a system-level
investigation of \gls{ul} access schemes under realistic conditions.

\section{System model}
\label{sec:system_model}

\begin{figure}[tb]
    \centering
    \scalebox{.95}{
    \begin{tikzpicture}[
        >={Stealth[length=4pt, width=3pt]},
        every node/.style={font=\small},
        state/.style={draw, circle, fill=white, font=\tiny,
                  inner sep=1pt, minimum width=1.1cm},
        ulink/.style={->, thick, blue!65!black},
        backhaul/.style={->, very thick, gray, dashed},
        framebox/.style={draw=black!50, fill=gray!4, rounded corners=2.5pt,
                     line width=0.5pt},
        iconUE/.style={font=\large, color=gray!75!black},
        iconENB/.style={font=\LARGE, color=black},
        iconPanel/.style={font=\normalsize, color=blue!60!black},
        iconServer/.style={font=\Large, color=blue!50!black},
        iconApp/.style={font=\LARGE, color=gray!50!black},
        iconSun/.style={font=\Large, color=yellow!75!orange},
        iconBat/.style={font=\normalsize, color=green!65!black},
        energyflow/.style={->, color=orange!75!black, line width=0.6pt,
                     line cap=round, >=Stealth},
                     ]

    \begin{scope}[shift={(0,5)}]
        \draw[framebox] (-0.12,-1.0) rectangle (3.3,1.0);
        \node[state] (inact) at (0.6,0.25)  {INACTIVE};
        \node[state] (act)   at (2.6,0.25)  {ACTIVE};
        \draw[->] (inact) to[bend left=22]
         node[above, font=\tiny] {$p_{IA}$} (act);
        \draw[->] (act)   to[bend left=22]
         node[below, font=\tiny] {$p_{AI}$} (inact);
        \node[font=\scriptsize, anchor=north] at (1.7,-0.45)
         {State evaluated every $T$s};
    \end{scope}

    \begin{scope}[shift={(4.15, 4.75)}]
        \draw[framebox] (-0.15,-0.75) rectangle (3.0,1.35);
        \node[iconPanel] (panel) at (0.55,0.8) {\faSolarPanel};
        \node[iconSun] (sun) at (panel.north west) {\faIcon[duotone]{sun-bright}};
        \node[font=\tiny, anchor=north] at (panel.south) {$P_h(t)$};
        \node[iconBat, scale=1.2] (b) at (2.1, 0.825)
            {\rotatebox{90}{\faIcon[duotone]{battery-three-quarters}}};
        \node[font=\tiny, anchor=north] at (b.south) {$C,\,v(t)$};
        \draw[->, color=orange!75!black, line width=0.6pt] (panel) -- (b);
        \node[font=\scriptsize, anchor=north west, align=left] at (-0.05,0.05)
            {Brown-out: $v(t) < v_\mathrm{th}$\\
             Operational: $v(t) > v_\mathrm{th}$};
    \end{scope}

    \begin{scope}[shift={(1.95,1.7)}]
        \draw[gray!50, dashed, line width=0.4pt] (0,0) circle (1.90cm);
        \node[iconENB] (enb) at (0,0) {\faIcon[duotone]{tower-cell}};

        \foreach \x/\y in {-1.65/0.42, -1.30/-1.05, -0.25/1.35,
                            0.65/-1.05, 0.95/0.75} {
        \node[iconUE, scale=1.1] (u) at (\x,\y) {\faIcon[duotone]{mobile-retro}};
        \node[iconPanel, scale=0.8] (p) at ($ (u.east) + (0.18, 0.10) $)
            {\faSolarPanel};
        \node[iconBat, scale=0.8] (b) at ($ (p.south) + (0, -0.16) $)
            {\rotatebox{90}{\faIcon[duotone]{battery-three-quarters}}};
        \draw[energyflow]
            ($(p.south) + (0, 0.16)$) to[out=-70, in=-50, looseness=3]
            ($(u.south) + (0.0, 0.25)$);
        \draw[ulink, dashed, line width=0.4pt] (u) -- (enb);
        }
    \end{scope}

    \begin{scope}[shift={(4.85, 2.6)}]
        \draw[framebox] (-0.2,-0.3) rectangle (2.3,0.8);
        \node[iconServer] (sgw) at (0.2,0.4) {\faServer};
        \node[iconServer] (mme) at (1.05,0.4) {\faServer};
        \node[font=\tiny, anchor=north] at (mme.south) {Core Network};
        \node[iconServer] (pgw) at (1.9,0.4) {\faServer};
    \end{scope}

    \node[iconApp] (app) at (6.15, 0.6) {\faDesktop};
    \node[font=\tiny, anchor=north] at (app.south) {App server};

    \draw[backhaul] (1.95,1.7) -- (4.85, 2.6);
    \draw[backhaul] (6.15, 2.28) -- ($ (app.north) + (0, 0.05) $)
            node[midway, right, font=\tiny] {$\tau_{p2p}$};

    \end{tikzpicture} }
    \vspace{-.2cm}
    \caption{System model overview: two-state Markov-modulated per-UE traffic
    source (top-left); per-UE energy front-end with ambient harvester $P_h(t)$
    feeding a leaky supercapacitor $(C, v(t))$ under voltage-threshold
    brown-out gating (top-right); single-cell deployment of
    \gls{aiot} \glspl{ue} (dashed blue: \gls{ul}) around the \gls{bs},
    connected to a remote application server through the core network via a
    fixed-delay backhaul (dashed gray).}
    \label{fig:system_model}
    \vspace{-.6cm}
\end{figure}
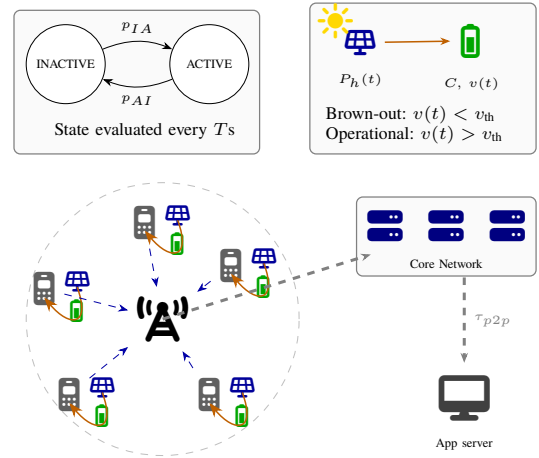

We consider a single-cell \gls{nb-iot} \gls{ul} scenario in which a population
of \gls{aiot} \glspl{ue} periodically reports short status messages to a remote
application server. We assume that the \glspl{ue} are immobile and uniformly
distributed around the \gls{bs} within a cell of radius $R~\text{meters}$. The
remote host is connected to the core network via a high-capacity point-to-point
link.

The traffic generated by each \gls{ue} is modeled as a two-state
Markov-modulated source $\mathcal{S} = \{\textsc{INACTIVE}, \textsc{ACTIVE}\}$,
with state transitions occurring every $T~\text{seconds}$. The transition
probabilities are denoted by $p_{\mathrm{IA}}$ for the transition from
\textsc{INACTIVE} to \textsc{ACTIVE}, and by $p_{\mathrm{AI}}$ for the reverse
transition. Therefore, the steady-state probability of being in the
\textsc{ACTIVE} state is $\pi_a = \nicefrac{p_{IA}}{(p_{IA} + p_{AI})}$. A
\gls{ue} generates a \gls{udp} packet of size $L~\text{bytes}$ at a decision
epoch whenever it is in the \textsc{ACTIVE} state. Accordingly, the long-term
mean offered traffic rate is $\nicefrac{8 \pi_a L}{T}\, \si{\bit\per\second}$
per \gls{ue}.

Additionally, each \gls{ue} is equipped with an energy-harvesting module
comprising a supercapacitor replenished by a time-varying solar energy source.
The harvested input power follows a diurnal $\sin^2$ profile based on the
photovoltaic power-system model of~\cite{Rekioua_2012},
\begin{equation}
  P_h(t) =
  P_h^{\max}\sin^{2}\!\left(\pi\!\left(t/T_\mathrm{day}+\phi\right)\right),
  \label{eq:solar_profile}
\end{equation}
where $P_h^{\max}$ denotes the peak harvested power,
$T_\mathrm{day}$ the harvesting period, and $\phi$ a configurable phase offset.
The resulting profile has an average harvested power equal to half of
$P_h^{\max}$.

The supercapacitor, with capacitance $C$, is modeled using the single-branch
equivalent circuit of~\cite{Spyker2000}, comprising an ideal capacitor in series
with an equivalent series resistance $R_\mathrm{s}$ and a parallel leakage
resistance $R_\mathrm{leak}$ accounting for self-discharge. The radio is
represented as a current sink rather than a resistive load, with
$i_\mathrm{load}$ obtained from the measured power of the active radio state, so
that consumption and self-discharge enter the model through separate terms. The
capacitor state is updated in an event-driven manner at every radio power-state
transition, harvest update, and at least every $\Delta t_\mathrm{poll}$,
assuming a piecewise-constant net current over each update interval. Given the
net current $i_k=i_\mathrm{h}(t_k)-i_\mathrm{load}(t_k)$ over an interval of
duration $\Delta t$, the capacitor voltage evolves according to the closed-form
solution of the first-order circuit,
\begin{equation}
  v(t_k+\Delta t)=
  R_\mathrm{eq}i_k
  \left(1-e^{-\Delta t/\tau}\right)
  +v(t_k)e^{-\Delta t/\tau},
  \label{eq:cap_evolution}
\end{equation}
where $\tau=R_\mathrm{eq}C$ and $R_\mathrm{eq}$ is dominated by $R_\mathrm{leak}$
because $R_\mathrm{leak}\gg R_\mathrm{s}$. Since every update interval satisfies
$\Delta t\ll\tau$, \eqref{eq:cap_evolution} reduces to
$\Delta v\approx i_k\Delta t/C-v(t_k)\Delta t/\tau$, that is, charge accounting
for the net current plus a slow self-discharge term. Energy drawn by the radio
is therefore carried by $i_\mathrm{load}$ rather than by the time constant.

The voltage is clamped to $[0,v_\mathrm{max}]$, and the stored energy is
$E(t)=\tfrac{1}{2}Cv^2(t)$. Device operation is governed by a
voltage threshold $v_\mathrm{th}$. When $v(t) < v_\mathrm{th}$, the \gls{ue}
enters a brown-out state in which radio transmission and protocol operations are
suspended and no packets are generated. The connection context is retained, so
recovery does not require a new attachment. Operation resumes once the voltage
exceeds $v_\mathrm{rec} = v_\mathrm{th} + \Delta v$, where $\Delta v$ introduces
hysteresis to avoid oscillatory switching near the threshold. This energy gating
couples the harvesting process with both traffic generation and radio-access
activity.

A packet is thus transmitted only when the \gls{ue} is in the \textsc{ACTIVE}
traffic state and not in brown-out, and an uplink grant has been obtained
through the configured access scheme. The energy consumed is determined by the
\gls{ue} radio state, with each state in Table~\ref{tab:nbiot_power_model}
assigned a constant power draw derived from practical \gls{nb-iot}
measurements~\cite{quectelBG96}. The system model, coupling stochastic traffic,
harvesting dynamics, and radio access, is illustrated
in~\Cref{fig:system_model}.

\section{Extension of the ns-3 NB-IoT Module}
\label{sec:impl_summary}

We extend the LENA-NB ns-3 platform~\cite{Jorke2022} with the components
required to support active \gls{aiot} operation together with considered
\gls{ul} access schemes. The resulting framework augments the \gls{nb-iot}
protocol stack with energy-constrained traffic generation, supercapacitor-based
energy storage, \gls{sr}, \gls{hsr} and \gls{fug}.

To model the intermittent traffic generated by \gls{aiot} devices,
we introduced a new application-layer class, \texttt{MarkovUdpClient}. This
class implements the two-state Markov-modulated \gls{udp} source described in
\ref{sec:system_model}. Energy storage is modeled through a new
\texttt{GenericCapacitor} class derived from \texttt{EnergySource}. The
implementation captures RC supercapacitor dynamics, including leakage effects,
internal resistance, voltage clamping at $v_\mathrm{max}$, and closed-form
capacitor-state evolution. Harvested energy is supplied by the
\texttt{SolarEnergyHarvester} class derived from \texttt{EnergyHarvester}, which
injects the diurnal power profile $P_h(t)$ defined in \eqref{eq:solar_profile}
into the capacitor. The \texttt{NbiotEnergyModel} class was extended with a
hysteretic brown-out mechanism and per-radio-state energy accounting. It also
exports a callback interface that notifies the \texttt{MarkovUdpClient}
application of brown-out events.

\begin{figure}[tb]
    \centering
    \scalebox{.85}{
    \begin{tikzpicture}[
        font=\footnotesize,
        >=Latex,
        node distance=0.5cm and 0.5cm,                  
        cls/.style    = {draw, rounded corners=2pt, align=center,
                     inner sep=3pt, minimum width=1.3cm, minimum height=0.7cm},
        nas/.style    = {cls, fill=blue!10, draw=black, thin},
        lte/.style    = {cls, fill=blue!10, draw=black, thin},
        nbiot/.style  = {cls, fill=blue!10, draw=black, thin},
        newcls/.style = {cls, fill=yellow!45, draw=orange!75!black, very thick,
                        double, double distance=0.7pt},
        modcls/.style = {cls, fill=orange!22,    draw=orange!75!black, very thick,
                        dashed, dash pattern=on 3.5pt off 2pt},
        zonebox/.style= {draw, dashed, rounded corners=3pt, gray, inner sep=8pt, fill=gray!3},
        has/.style    = {-Latex, thick},
        sap/.style    = {-Latex, dashed, thick},
        tr/.style     = {-Latex, densely dotted, thick, color=blue!70!black},
    ]

    \node[newcls] (markov)                       {MarkovUdpClient};
    \node[nas, minimum width=5.0cm]    (nas)    [below=of markov]
        {EpcUeNas/EpcEnbNas};
    \node[modcls, minimum width=5.0cm] (rrc)    [below=of nas]
        {LteUeRrc/LteEnbRrc};
    \node[lte]    (rrcp)   [below=of rrc]         {LteUe/EnbRrcProtocol};

    \node[nas]    (pdcp)   [below=of rrcp, xshift=-0.7cm]       {LtePdcp};

    \node[lte]    (rlcam)  [below=of pdcp]      {LteRlcAm};
    \node[lte]    (rlctm)  [right=of rlcam, xshift=-0.4cm]      {LteRlcTm};
    
    \node[lte, modcls, minimum width=4.0cm]    (mac)    [below=of rlcam,
        xshift=.7cm]       {LteUeMac/LteEnbMac};
    \node[lte, modcls, minimum width=5.0cm]    (phy)    [below=of mac]
        {LteUePhy/LteEnbPhy};

    \node[nbiot, minimum width=1.5cm]  (amc)    [right=of rrc, xshift=0.6cm, yshift=0.6cm]         {NbiotAmc};
    \node[modcls] (sched)  [below=of amc]         {NbiotScheduler};

    \node[newcls] (harv)   [below=1.03cm of sched] {SolarEnergyHarvester};
    \node[newcls, minimum width=2.6cm] (cap)    [below=of harv]        {GenericCapacitor};
    \node[modcls] (egy)    [below=of cap]         {NbiotEnergyModel};

    \draw[has, transform canvas={xshift=0.5cm}]  (markov) -- (nas);
    \draw[has, transform canvas={xshift=-.5cm}] (nas) -- (markov);
    \draw[has, transform canvas={xshift=0.9cm}]  (nas) -- (rrc);
    \draw[has, transform canvas={xshift=-.9cm}] (rrc) -- (nas);
    \draw[has, transform canvas={xshift=0.7cm}]  (rrc) -- (rrcp);
    \draw[has, transform canvas={xshift=-.7cm}] (rrcp) -- (rrc);

    \draw[has] ([xshift=-0.4cm]rrcp.south) -- ([xshift=0.3cm]pdcp.north);
    \draw[has] ([xshift=-0.3cm]pdcp.north)  -- ([xshift=-1cm]rrcp.south);

    \draw[has, transform canvas={xshift=0.3cm}]  (pdcp) -- (rlcam);
    \draw[has, transform canvas={xshift=-0.3cm}] (rlcam) -- (pdcp);

    \draw[has] ([xshift=0.5cm]rrcp.south) -- ([xshift=-0.3cm]rlctm.north);
    \draw[has] ([xshift=0.3cm]rlctm.north)  -- ([xshift=1.1cm]rrcp.south);

    \draw[has] ([xshift=-1.0cm]mac.north) -- ([xshift=-0.3cm]rlcam.south);
    \draw[has] ([xshift=0.3cm]rlcam.south)  -- ([xshift=-0.4cm]mac.north);

    \draw[has] ([xshift=1.1cm]mac.north) -- ([xshift=0.3cm]rlctm.south);
    \draw[has] ([xshift=-0.30cm]rlctm.south)  -- ([xshift=0.5cm]mac.north);

    \draw[has, transform canvas={xshift=-.7cm}] (mac) -- (phy);
    \draw[has, transform canvas={xshift=0.7cm}]  (phy) -- (mac);

    \draw[has] ([xshift=-1.7cm]mac.north) -- ([xshift=-1.7cm]rrc.south);
    \draw[has] ([xshift=1.7cm]rrc.south)  -- ([xshift=1.7cm]mac.north);

    \draw[has] ([xshift=-2.15cm]phy.north) -- ([xshift=-2.15cm]rrc.south);
    \draw[has] ([xshift=-2.4cm]rrc.south)  -- ([xshift=-2.4cm]phy.north);

    \draw[has] (sched.north) -- (amc.south);

    \draw[sap] (rrc.east) (rrc.east) -| ++(0.2,0) |-   ([yshift=-0.2cm]egy.west);
    \draw[sap] ([yshift=-0.2cm]mac.east) --   ([yshift=-0.2cm]egy.west);
    \draw[sap] (phy.east) -|   (egy.south);

    \draw[has] ([yshift=0.2cm]mac.east) -- ++(0.3,0) |- (sched.west);

    \draw[sap]  (cap)     -- (harv)
         node[midway,right,font=\scriptsize, color=black]{harvest};
    \draw[sap] (egy.north) -- (cap.south)
         node[midway,right,font=\scriptsize, color=black]{drain (gated)};


    \draw[sap] (egy.east) (egy.east) -| ++(0.5,0) |- (markov.east)
        node[ pos=0.75, above, font=\scriptsize, color=black]
          {brown-out callback (Pause/Resume)};


    \draw[gray!50, thin]
    ([yshift=-0.4cm]phy.south west) -- ++(8cm, 0);

    \begin{scope}[
        shift={([yshift=-0.7cm]phy.south west)},
        every node/.append style={font=\scriptsize}
    ]
    \node[newcls, minimum width=0.55cm, minimum height=0.35cm,
          inner sep=1pt, xshift=0.7cm]                                    (s_new)   {};
    \node[anchor=west, right=2pt of s_new]                  (l_new)   {Newly added};

    \node[modcls, minimum width=0.55cm, minimum height=0.35cm,
          inner sep=1pt, right=0.5cm of l_new]              (s_mod)   {};
    \node[anchor=west, right=2pt of s_mod]                  (l_mod)   {Modified};

    \node[nas, minimum width=0.55cm, minimum height=0.35cm,
          inner sep=1pt, right=0.5cm of l_mod]              (s_unmod) {};
    \node[anchor=west, right=2pt of s_unmod]                (l_unmod) {Unmodified};

    \draw[has] ([yshift=-0.55cm]s_new.west) -- ++(0.55cm, 0)
          coordinate (a_has_end);
    \node[anchor=west, right=2pt of a_has_end]              (l_has)   {NB-IoT-related call};

    \draw[sap] ([yshift=-0.55cm, xshift=0.55cm]s_mod.west) -- ++(0.55cm, 0)
          coordinate (a_sap_end);
    \node[anchor=west, right=2pt of a_sap_end]              (l_sap)   {Energy-related call};
  \end{scope}

\end{tikzpicture}}
\caption{Abridged ns-3 NB-IoT models extended for ambient IoT.}
\label{fig:class_diagram}
\vspace{-0.6cm}
\end{figure}
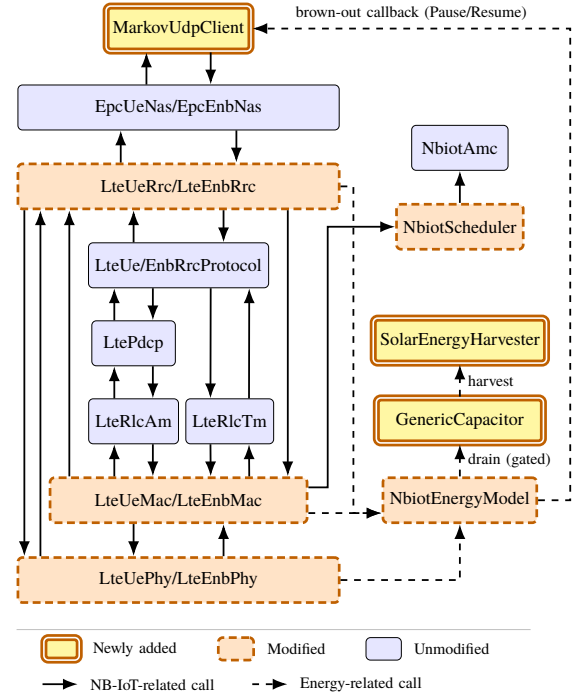

To support persistent \gls{ul} access and \gls{fug}, the \gls{rrc}
and \gls{mac} layers were extended to retain the \gls{ue} context across sleep
intervals. The \texttt{LteEnbRrc} and \texttt{LteUeRrc} classes implement
configurable massive-\gls{iot} timers (i.e., $T_{3324}$, $T_{3412}$,
$T_\mathrm{eDRX}$), context retention, suspend/resume procedures and
\gls{rai}-assisted connection release. Suspension is synchronized with the
\gls{rlc} acknowledgement state to ensure that pending acknowledgements are
exchanged before the connection is suspended. \texttt{LteUeMac} implements
dedicated \gls{sr} using reserved \gls{nprach} resources and opportunistic
contention on shared resources for \gls{hsr}. The corresponding \gls{bs}-side
resource mapping, contention resolution, and \gls{edt} procedures are
implemented in \texttt{LteEnbMac}, \texttt{LteEnbRrc}, and
\texttt{NbiotScheduler}. For proactive \gls{fug}, \texttt{NbiotScheduler}
schedules an uplink grant for each \gls{ue} at the predicted next transmission
time. The added and modified classes are summarized in
Fig.~\ref{fig:class_diagram}. The extended module, evaluation scenario, and
analysis scripts are publicly available on Github~\cite{agbeve2026ns3nbiot}.

\section{Performance Evaluations}
\label{sec:perf_eval}

\begin{table*}[!t]
    \centering
    \caption{Simulation parameters; per-state power draw from Quectel BG96
    measurements~\cite{quectelBG96}}
    \vspace{-0.2cm}
    \label{tab:sim_params}
    \label{tab:nbiot_power_model}
    \renewcommand{\arraystretch}{1.0}
    \footnotesize
    \begin{tabular}{lll|lll}
        \hline
        \textbf{Parameter} & \textbf{Symbol} & \textbf{Value} &
        \textbf{Parameter} & \textbf{Symbol} & \textbf{Value} \\
        \hline
        Cell radius                & $R$                 & \qty{1.25}{\kilo\meter} &
        DCI~0 scheduling delay     & $k_0$               & \qty{8}{\milli\second} \\
        Number of UEs              & $N_\mathrm{UE}$     &
                                     $\numrange[range-phrase = {-}]{1}{200}$ &
        Contention resolution      & $T_\mathrm{CR}$     & \qty{512}{\milli\second} \\
        UE / BS Tx power           & $P_t^{UE},P_t^{BS}$ &
                                                \qtylist[list-units=single,
                                    list-pair-separator={,~}]{10;30}{\dBm} &
        NPDCCH USS config          &
                            $R_\mathrm{max}^\mathrm{USS},\,G^\mathrm{USS}$ &
                        \numlist[list-pair-separator={,~}]{8;2}
                        ($T_\mathrm{USS}{=}$\qty{16}{\milli\second}) \\
        Noise figure               & $NF$                & \qty{0}{\decibel} &
        \glsxtrshort{psm} power draw   & $P_\mathrm{PSM}$    & $\qty{15}{\micro\watt}$ \\
        Pathloss model             & --                  & Friis &
        \glsxtrshort{edrx} / connected-idle power draw & $P_\mathrm{idle}$ &
                                                   $\qty{3}{\milli\watt}$ \\
        P2P link delay             & $\tau_\mathrm{p2p}$ & \qty{10}{\milli\second} &
        \glsxtrshort{npdcch} / \glsxtrshort{npdsch} Rx power draw & $P_\mathrm{rx}$ &
                                                  $\qty{80}{\milli\watt}$ \\
        Application payload        & $L$                 & \qty{49}{\byte} &
        \glsxtrshort{nprach} / \glsxtrshort{npusch} Tx power draw & $P_\mathrm{tx}$ &
                                                 $\qty{589}{\milli\watt}$ \\
        State update interval      & $T$                 & \qty{300}{\second} &
        Capacitance                & $C$                 & \qty{1}{\farad} \\
        Transition probabilities   & $p_{IA},p_{AI}$     &
                                \numlist[list-pair-separator={,~}]{0.7;0.2} &
        Initial / Max voltage                & $v_0, v_\mathrm{max}$
                                             &$\sim\mathcal{U}(\numlist[list-pair-separator={,~}]{2.9;3.3})$, \qty{3.3}{\volt} \\
        RLC mode                   & --                  & AM &
        Cut-off / recovery voltage & $v_\mathrm{th},v_\mathrm{rec}$ &
                                                    \qtylist[list-units=single,
                                list-pair-separator={,~}]{2.4;2.6}{\volt} \\
        Early Data Transmission    & --                  & enabled &
        Internal / leakage $R$     & $R_\mathrm{s},R_\mathrm{leak}$ &
                                                \qtylist[list-units=single,
                                    list-pair-separator={,~}]{0.1; e4}{\ohm} \\
        NPRACH periodicity (\glsxtrshort{ce}-0) & $T_\mathrm{NPRACH}$ &
                                                    \qty{80}{\milli\second} &
        Harvest profile            & $P_h(t)$            &
                                    $\sin^{2}$, eq.~\eqref{eq:solar_profile} \\
        NPRACH / reserved \gls{sr} subcarriers &
                            $P_\mathrm{sc},P_\mathrm{sc}^\mathrm{SR}$ &
                                \numlist[list-pair-separator={,~}]{12;6} &
        Peak harvest power  & $P_h^\mathrm{max}$ &
                                            \qtylist[list-units=single,
                                            list-pair-separator={,~},
                            list-final-separator={,~}]{1; 1.5; 2}{\milli\watt} \\
        \glsxtrshort{sr} reservation validity & $T_\mathrm{res}$ &
                                       \qty{3}{\hour} &
        Energy poll interval       & $\Delta t_\mathrm{poll}$ & \qty{100}{\milli\second} \\
        NPRACH reps / max attempts & $N_\mathrm{rep}$, -- &
                                \numlist[list-pair-separator={,~}]{1;10} &
        Active / \glsxtrshort{tau} timer & $T_{3324},T_{3412}$ &
                                       \qty{2}{\second}, \qty{24}{\hour} \\
        RAR window                 & $W_\mathrm{RAR}$    & \qty{160}{\milli\second} &
        eDRX cycle length          & $T_{\mathrm{eDRX}}$ & \qty{40.96}{\second} \\
        MSG3 scheduling delay      & $k_2$               & \qty{12}{\milli\second} &
        Data-inactivity release    & $T_\mathrm{rel}$    & \qty{1}{\second} \\
        NPUSCH duration            & $T_\mathrm{NPUSCH}$ & \qty{4}{\milli\second} &
        Simulation duration        & $T_\mathrm{sim} = T_\mathrm{day}$    & \qty{1}{\hour} \\
        \hline
    \end{tabular}
    \vspace{-.7cm}
\end{table*}

\subsection{Simulation setup}
\label{subsec:simsetup}

\begin{figure*}[tb]
  \centering
  \subfloat[\label{fig:delay}]{%
    \includegraphics[width=0.32\textwidth, trim=0 6 0 0]{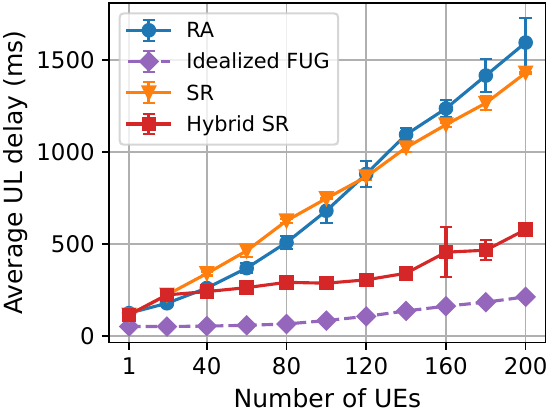}
    }
  \hfill
  \subfloat[\label{fig:eperbit}]{%
    \includegraphics[width=0.32\textwidth, trim=0 6 0 0]{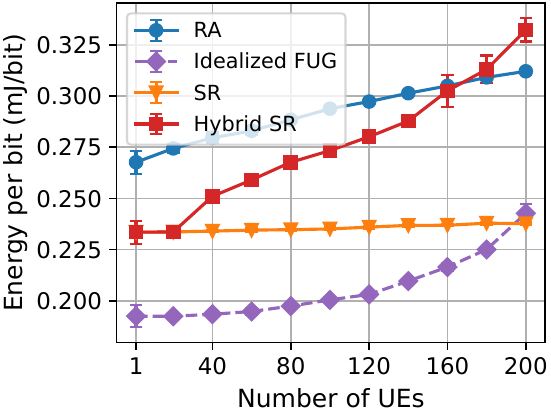}
    }
  \hfill
  \subfloat[\label{fig:depleted}]{%
    \includegraphics[width=0.32\textwidth,trim=0 6 0
    0]{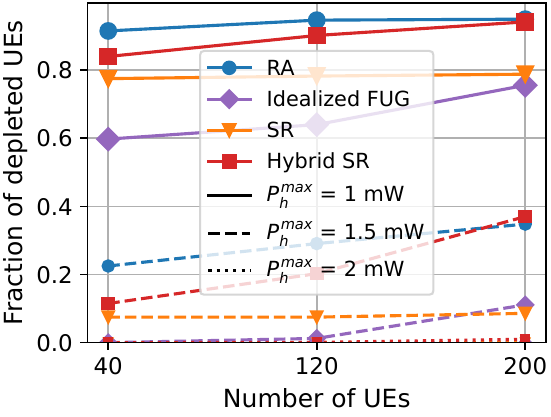}
    }
  \caption{Performance vs.\ the number of \glspl{ue}
      ($P_h^\mathrm{max}{=}\qty{2}{\milli\watt}$ for (a), (b)):
    (a) mean uplink delay $\bar{D}$,
    (b) energy per delivered bit $\varepsilon$.
    (c) Fraction of \glspl{ue} with at least one brown-out}
  \label{fig:results_main}
  \vspace{-0.3cm}
\end{figure*}

\Glspl{ue} start pre-registered but suspended in an idle state, such that the
first \gls{ul} packet triggers the standard \gls{nb-iot} \gls{ra} procedure and
establishes the connection context. Connected \glspl{ue} are released after a
data-inactivity timeout of $T_{\mathrm{rel}}=\qty{1}{\second}$ and transition
through the \gls{aiot} timer profile (i.e., $T_{3324}{=}2$~\si{\second},
$T_{\mathrm{eDRX}}{=}\qty{40.96}{\second}$, and $T_{3412}{=}\qty{24}{\hour}$)
into \gls{psm}. This life cycle is common to all schemes, including Idealized
\gls{fug}. For the later, each packet arrival is known to the \gls{bs} instantly
through a \gls{bsr} of zero latency and signaling cost. Since a packet arrival
coincides with the \gls{ue} wake-up, the \gls{bs} schedules the \gls{ul} grant
at the next available \gls{npdcch} opportunity. \gls{edt} is enabled throughout,
allowing each contention-based access to carry the application payload in Msg3.
Traffic follows the Markov source of Sec.~\ref{sec:system_model} with
$T=\qty{300}{\second}$, $(p_{IA},p_{AI})=(0.7,0.2)$, $L=\qty{49}{\byte}$.

Protocol processing follows the default \gls{nb-iot} stack, with \gls{pdcp} and
\gls{rlc} operating in acknowledged mode and without header compression.
\gls{nprach} opportunities recur every $\qty{80}{\milli\second}$, while the
scheduler allocates resources at each \gls{uss} occasion
($T_{\mathrm{USS}}{=}\qty{16}{\milli\second}$), using a round-robin policy. All
\gls{ul} transmissions employ single-tone $\qty{15}{\kilo\hertz}$ \gls{npusch}
Format~1. Under \gls{sr} and \gls{hsr}, six of the twelve \gls{ce}-\num{0}
\gls{nprach} subcarriers are reserved for dedicated \gls{sr} preambles
($P_\mathrm{sc}^\mathrm{SR}{=}\num{6}$) while the remaining six are available
for contention-based access and the initial attachment procedure. Propagation is
modeled using the Friis path-loss model. Since the simulated cell dimension
($R{=}\qty{1.25}{\kilo\meter}$) is relatively small for \gls{nb-iot} networks under
free space propagation, all UEs operate under the same maximal \gls{mcs} and the
same \gls{ce}-\num{0}.
This isolates the comparison from link-adaptation
and coverage-extension effects that would otherwise depend on \gls{ue}
placement. 


Each \gls{ue} is powered by a $\qty{1}{\farad}$ supercapacitor
($v_{\max}=3.3$~V, $v_{\mathrm{th}}=2.4$~V) replenished by the diurnal
harvesting profile of \eqref{eq:solar_profile} with a peak harvested power
varied across \qtylist[list-units=single,list-pair-separator={,~}]{1; 1.5;
2}{\milli\watt}. The initial voltage is drawn independently per device from
$v_0\sim\mathcal{U}(\numlist[list-units=single,
list-pair-separator={,~}]{2.9;3.3})~\si{\volt}$ to avoid synchronized
cold-start.
A hysteretic brown-out mechanism with a recovery threshold of
$v_\mathrm{rec}=2.6$~V disables both radio activity and application traffic
until the capacitor voltage exceeds the recovery threshold. The per-state radio
power consumption follows the Quectel BG96 measurements. The complete set of
simulation parameters is summarized in Table~\ref{tab:sim_params}.

\subsection{Discussion of results}
\label{sec:results}

We evaluate the access schemes using network-level and energy-related metrics.
Each data point averages ten independent runs with different seeds; error bars
indicate standard deviation.
Reliability is measured by the mean \gls{ul} delay $\bar{D}$
over delivered packets. Energy efficiency is quantified by the energy per delivered
bit $\varepsilon=\nicefrac{E_{\mathrm{alive}}}{B_{\mathrm{rx}}}$, where
$E_{\mathrm{alive}}$ is the energy consumed by a \gls{ue} while
it is not brown-out and $B_{\mathrm{rx}}$ the delivered payload.
Sustainability is evaluated by the brown-out fraction $\phi$, the fraction of
devices whose capacitor voltage falls below $v_{\mathrm{th}}$ at least once.

\subsubsection{Network-level metrics}


\looseness=-1
\Cref{fig:delay} shows the corresponding mean \gls{ul} delay $\bar{D}$. The
Idealized \gls{fug} provides a lower bound, with delay increasing from
\qty{50}{\milli\second} to \qty{212}{\milli\second} as grant deliveries
serialize over shared \gls{npdcch} occasions at high density. Among
the deployable schemes, \gls{hsr} exhibits the slowest increase in delay as the
number of \glspl{ue} grows, from \qty{222}{\milli\second} at
$N_{\mathrm{UE}}=\num{20}$ to \qty{595}{\milli\second} at
$N_{\mathrm{UE}}=\num{200}$. This limited increase results from the
opportunistic contention path, which resolves most requests before the dedicated
request occasion is reached. In contrast, the mean delay of \gls{sr} increases
from \qty{222}{\milli\second} to \qty{1.43}{\second} as the round-robin request
period increases with the number of \glspl{ue}. The delay of \gls{ra} similarly
increases from \qty{177}{\milli\second} to \qty{1.60}{\second} as contention
intensifies. At $N_{\mathrm{UE}}=\num{200}$, \gls{hsr} therefore reduces the
mean \gls{ul} delay by a factor of $2.4$--$2.7$ relative to \gls{sr} and
\gls{ra}, respectively.

\subsubsection{Energy efficiency and sustainability}

\looseness=-1
\Cref{fig:eperbit} shows the energy per delivered bit $\varepsilon$.
Beyond $N_{\mathrm{UE}}\approx\num{180}$, idealized \gls{fug} becomes less
energy-efficient than \gls{sr}, as increasing grant serialization extends
\gls{npdcch} monitoring for waking \glspl{ue}.
\Gls{sr} is the most energy-efficient among the deployable schemes and remains
nearly insensitive to the number of \glspl{ue}, with $\varepsilon$ ranging from
\qty{0.234}{\milli\joule\per\bit} to \qty{0.238}{\milli\joule\per\bit}.
Dedicated, collision-free request opportunities keep the per-packet awake time
nearly constant, while the waiting time for the reserved occasion is spent in
\gls{psm}. \gls{ra} consumes \qtyrange[range-phrase={{-}},
range-units=single]{17}{31}{\percent} more energy per delivered bit than
\gls{sr}, increasing from \qty{0.27}{\milli\joule\per\bit} to
\qty{0.31}{\milli\joule\per\bit} as contention increases the time spent
accessing the channel. \Gls{hsr} has energy consumption close to \gls{sr} at low
density but increases to \qty{0.33}{\milli\joule\per\bit} at
$N_{\mathrm{UE}}=\num{200}$, exceeding \gls{ra} at approximately
$N_{\mathrm{UE}}=\num{170}$. At high density, three effects contribute to this
increase: nearly every packet contends before its reserved request occasion, the
contention request partition is half the size of the \gls{ra} pool, and each
failed attempt incurs a full response-window monitoring period. Thus,
\gls{hsr}'s lower request latency comes at the cost of increased contention
energy, whereas \gls{sr} maintains nearly constant energy consumption by trading
the same increase in density for request delay.

\Cref{fig:depleted} reports the brown-out fraction $\phi$. At a peak harvested
power of \qty{2}{\milli\watt}, all schemes complete every run without brown-out,
with $\phi\approx\num{0}$. At \qty{1.5}{\milli\watt}, differences in energy
consumption become apparent: \gls{sr} maintains a nearly load-independent
$\phi\approx\num{0.08}$, whereas \gls{ra} increases from \num{0.23} to
\num{0.35} and \gls{hsr} from \num{0.12} to \num{0.37}. The latter exceeds
\gls{ra} at high density, consistent with its higher contention energy. At
\qty{1}{\milli\watt}, none of the schemes can sustain operation, with $\phi$
ranging from \num{0.60} to \num{0.95}. Thus, brown-out becomes significant as
the harvested power decreases below the level required for sustainable
operation.

\section{Conclusion}
\label{sec:conclusion}

\looseness=-1
We studied the energy--latency trade-off of \gls{ul} access for energy-harvesting
\gls{aiot} devices by comparing \gls{ra}, \gls{sr}, the proposed
\gls{hsr}, and \gls{fug}. The Idealized \gls{fug} configuration establishes a
lower bound on achievable delay and energy cost when future traffic is known
perfectly, while practical \gls{fug} performance degrades with traffic
prediction errors. Among the implementable access schemes, \gls{sr} provides the
best energy efficiency, achieving approximately
\qty{0.24}{\milli\joule\per\bit}, \qtyrange[range-phrase={--}]{17}{31}{\percent}
below \gls{ra}. In contrast, \gls{hsr} provides the lowest delay, reducing mean
\gls{ul} delay by a factor of
$\numrange[range-phrase={{-}}]{2.4}{2.7}\times$ relative to \gls{sr} and
\gls{ra} at $N_{\mathrm{UE}}=\num{200}$, respectively. This delay reduction comes at the cost
of additional contention energy, causing \gls{hsr} to become less
energy-efficient than \gls{ra} beyond approximately \num{170} \glspl{ue}.
The preferred \gls{ul} access scheme therefore depends on network density,
latency requirements, and available harvested energy. Future work will develop
heuristic and learning-based predictive \gls{fug} schedulers that operate
without perfect traffic knowledge and evaluate how closely they can approach
the delay and energy lower bounds established by Idealized \gls{fug}.

\section*{Acknowledgement}
This work was funded by the AMBIENT-6G project of the Smart Networks and
Services Joint Undertaking (SNS JU), EU Horizon Europe Grant
No.~101192113.

\printbibliography[title={References}]
\end{document}